%% file: EfficientAlgorithm.tex
\documentclass[a4paper,11pt]{article}
\usepackage[a4paper, margin=22.5mm]{geometry}

\usepackage{amsmath}
  \numberwithin{equation}{section}
  \usepackage{amssymb}

\usepackage{enumerate}
\usepackage{algorithmic}
\usepackage{graphicx}

\usepackage{LifeCon}

\usepackage[round,sort]{natbib}
\usepackage{color}
\usepackage{ulem}

\begin{document}

\title{An efficient algorithm for the calculation of reserves for non-unit linked life policies}
\author{Mark Tucker\textsuperscript{*} and J. Mark Bull \\ Edinburgh Parallel Computing Centre}
\date{}
\maketitle

\input{abstract}


{\small  {\bf Correspondence to}: Mark Tucker, EPCC, The University of Edinburgh, King's Buildings, Mayfield Road, Edinburgh, EH9 3JZ, UK. email: M.Tucker@epcc.ed.ac.uk}

\input{introduction}
\input{motivation}

\input{derivation}

\input{useCases}
\input{performance}

\input{conclusion}

\input{acknowledge}

\bibliography{references}

\include{representativePortfolio}
\include{runtimes}
\include{plot1}
\include{plot2}

\end{document}

%% file: abstract.tex
\begin{abstract}
The underlying stochastic nature of the requirements for the Solvency II regulations has introduced significant challenges if the required calculations are to be performed correctly, without resorting to excessive approximations, within practical timescales. It is generally acknowledged by practising actuaries within UK life offices that it is currently impossible to correctly fulfil the requirements imposed by Solvency II using existing computational techniques based on commercially available valuation packages.

Our work has already shown that it is possible to perform profitability calculations at a far higher rate than is achievable using commercial packages. One of the key factors in achieving these gains is to calculate reserves using recurrence relations that scale linearly with the number of time steps.

Here, we present a general vector recurrence relation which can be used for a wide range of non-unit linked policies that are covered by Solvency II; such contracts include annuities, term assurances, and endowments. Our results suggest that by using an optimised parallel implementation of this algorithm, on an affordable hardware platform, it is possible to perform the `brute force' approach to demonstrating solvency in a realistic timescale (of the order of a few hours).
\end{abstract}

%% file: introduction.tex
\section{Introduction}
  \label{Introduction}

In the UK, life companies are required to demonstrate solvency on a regular basis; the calculations for this are usually done using software created using commercial packages, such as Prophet, MoSes, or Algo Financial Modeller.  The requirements for Solvency II, if implemented naively, lead to the need for a far greater volume of calculations.  This volume of calculations is so vast that the use of commercially available software to obtain the results is currently beyond contemplation.

Many of the currently available packages produce single-threaded executables although, in the last few years, some packages have begun to venture into the world of multi-threaded programs.  Further, some of the commercial packages use naive summations to calculate reserves, while others use an approach based on commutation functions.  Typical implementations using either summations or commutation functions lead to computational complexity which is quadratic in the number of time steps; \cite{Macdonald} provides an overview of commutation functions from which it is possible to deduce that commutation functions lead to quadratic computational complexity.  The combination of single-threadedness and inefficient methodologies means that the performance of the programs produced by these packages is far below that required for demonstrations of solvency under the new regulations.

Over the last two decades, conventional wisdom within life offices has been to improve computing performance by obtaining more up-to-date PCs; it has been sufficient to rely on chip manufacturers increasing processing power of computers roughly in line with Moore's law.  However, physical considerations dictate that it is now no longer possible to rely on the speed of CPUs increasing in order to improve the processing power of computers.  Instead, over the past few years, chip manufacturers have moved to placing more compute cores on each chip in order to increase their processing power.  This means that, to benefit from modern processors, it is necessary to embrace programming paradigms which harness the power of the multi-core chips.  Whilst some commercial valuation packages do now have some parallel computation capability, these capabilities are often limited.

A reasonably common alternative approach to achieving higher throughput has been to split the data and calculations across large numbers of PCs to perform the calculations, and then combine the results when the calculations are finished.  The run times involved mean that using this approach to produce solvency results from, say, 1000 simulations at each future time step is not practical.

Using recurrence relations instead of summations or commutation functions leads to computational complexity which is linear in the number of time steps; for a typical projection, using monthly steps over 25 to 50 years, the saving is significant -- around two orders of magnitude.

Our previous results \citep{TuckerBull_WHPCF} show that, using a 48-core Symmetric Multi-Processor (SMP), it is possible to perform profitability calculations for single-life annuities, using monthly steps, at a rate of $3.9 \times 10^5$ policies per second; the commercial package we used as a comparison processes these policies at 1 per second.  The five orders of magnitude improvement in performance comes from a combination of factors; two orders from a change of hardware and parallelisation of the code using OpenMP \citep{OpenMP_3.1} to coordinate 48 threads, two orders from changing to use the recurrence relation, and the last order from further manual optimisation of the code.

Here we present a formulation of the recurrence approach in vector form, demonstrate how this relates to a variety of non-unit linked policies, and state the assumptions which are required in order that the recurrence relation holds.  We provide motivation using annuities, since these are the contracts which prompted this investigation; the vast numbers of policies and the associated reserves make these policies of prime importance to practising actuaries.  We develop the theory for a very general case which can be used for a wide range of life policies, and then provide specific in-depth examples for various policy types.  Finally, we present some performance results which demonstrate that an efficient implementation of the recurrence relation can lead to run times which make the Brute Force approach to calculating solvency computationally tractable.

%% file: motivation.tex
\section{Motivation}
  \label{Motivation}

Any company competing in the life assurance and pensions markets in the UK must demonstrate solvency as part of the regulations.  Throughout Europe, the Solvency II project \citep{Solvency_II} is introducing tougher requirements on both the capital requirement (i.e. the demonstration of solvency) and risk management aspects of the insurance industry; our work concentrates on the calculation of liabilities, which is a major part of the calculation of the capital requirement.

The existing regulations require that a single, `best estimate', calculation of solvency is performed using a set of assumptions chosen by actuaries.  However, using industry-standard software, the processing for a typical portfolio of annuity policies takes one-to-two days to complete on a single modern PC.  Under Solvency II, it is necessary for an insurer to hold sufficient capital such that there is only a 1-in-200 chance of being unable to cover the liabilities.

Therefore, one widely-adopted approach to satisfying the new regulations is to replace the single calculation of the liabilities by the generation of 1000 Monte Carlo scenarios at each future time step; projections typically use monthly time steps and extend for about 60 years.  From these 1000 scenarios generated, we seek the one which produces the 5\textsuperscript{th} largest liability value since that corresponds to a 0.5\% chance of the actual liabilities being larger than the value reported.  This is computationally intractable using current commercial software and so it is necessary to seek efficient algorithms, and implement those algorithms on hardware capable of highly parallel threading.

A significant amount of work has already been done on changing the way in which the stochastic processes underlying the assets or liabilities are modelled, with those changed processes being implemented on small scale parallel computers; see, for example, \cite{Corsaro_2010} where the focus is on the development of a parallel algorithm for the valuation of profit sharing policies.  In our work, we approach the problem differently; the generation of 1000 Monte Carlo scenarios at each future time step has become colloquially known in the UK as the ``Brute Force'' approach.  We aim to show that it is possible to perform the full brute force calculation (with a full valuation for each policy, for each scenario, for each step) within practical timescales using massively parallel architectures.  In order to achieve this, a fundamentally different approach to the calculations is required; we present an algorithm which is linear rather than quadratic in time steps and hence reduces the overall time required to perform the brute force calculations by a significant factor.

\subsection{The Brute Force Approach}
     \label{BruteForceApproach}

Within industry in the UK, current standard practice is that the calculations to value the liabilities are performed using programs created on specialist software packages which are crafted for flexibility and ease of use, rather than performance of the resulting programs.  These packages require the relationships between variables to be entered in a pseudo-code style language, and the package then produces source code, compiles and links it, and launches the resulting executable.  The clear advantage of these packages is that quite complex programs can be developed by users who are not trained programmers, allowing users who understand the intricacies of the policies to produce the programs; this is clearly beneficial when complicated contracts are being modelled.

The fact that the users are usually not trained programmers is also one of the main disadvantages of these packages; the art of programming is removed by the package and this usually results in the executable having sub-optimal performance. It is currently not uncommon for life assurance offices to have hundreds of PCs as dedicated slaves to perform these calculations; the problem with this approach lies in the fact that once these machines have completed their calculations, the results need to be collated and that renders this approach impractical.

It is generally acknowledged by practising actuaries that it is impossible to comply with the incoming regulations using the current approach. The policies which have been chosen to base this project on are life annuities; the payments made to a pensioner after retirement form a typical life annuity.  The profitability calculations required for these policies, using the packages outlined above and hardware which the industry is comfortable with, currently run at a rate of approximately one second per policy; it takes roughly 35 hours to produce one set of best estimate results for one block of 160,000 policies.

Performing 1000 simulations at each future time step using this technology is infeasible; supposing that, on average, each policy has 60 years (i.e. 720 months) outstanding at the valuation date, the estimated run time for this approach is
\begin{align*}
        1000 \times  \left[ \frac{720}{720} + \frac{719}{720} + \frac{718}{720} + \cdots + \frac{1}{720} \right] \times 35 \text{ hours}
     &  \approx 1.26 \times 10^7 \text{ hours}
  \\ &  \approx 1440 \text{ years.}
\end{align*}
Clearly, even with thousands of PCs, this is beyond contemplation.

Our previous work \citep{TuckerBull_WHPCF} shows that, by using straightforward optimisation techniques, it is possible to obtain significantly better performance than can be obtained from the commercial packages.  Firstly, using a good set of compiler options (including disabling debugging information) gave a speedup of $3.8\times$.  Secondly, manually optimising dominant routines led to a further speedup of about $12.4\times$; these changes included unrolling nested loops, simplifying arithmetical steps, removing initialisation of arrays which hold results, and replacing power calculations with successive multiplications.  Finally, changing to more modern hardware led to a speedup of about $1.87\times$ so that, combining all of these optimisations, the overall improvement was a speedup of about $88\times$.  Even using this optimised code, the estimated run time for the brute force approach is roughly 16 years, and this is still beyond contemplation. Using the recurrence approach is critical in order to reduce this to practical timescales.

\subsection{Actuarial Notation}
     \label{ActuarialNotation}

As far as reasonably practical, we use standard International Actuarial Notation \citep{GreenTables}, so that
\begin{align*}
    {}_{t}p_{x} &= \text{Pr[life aged precisely $x$ survives until age $x+t$]}
\intertext{and}
    {}_{t}q_{x} &= \text{Pr[life aged precisely $x$ dies before reaching age $x+t$]}
\end{align*}
As usual, if $t=1$, then the prefix is dropped so that $p_{x} = {}_{1}p_{x}$ and $q_{x} = {}_{1}q_{x}$.  Also following standard notation, for $t \ge 0$, $l_{x+t}$ is the number of lives expected to be alive at age $x+t$ given that there are $l_{x}$ lives at age $x$.

Sections \ref{Annuities} and \ref{VectorRecurrence} consider annuities; standard notation uses
\begin{itemize}
  \item[-] $a_x$ to represent the expected present value of an annuity where payments of amount 1 are made at the {\it end} of a year to a life aged $x$ at the time of valuation, so long as the life is alive at the time of payment, and
  \item[-] $\ddot{a}_x$ to represent the expected present value of an annuity where payments of amount 1 are made at the {\it start} of a year to a life aged $x$ at the time of valuation, so long as the life is alive at the time of payment.
\end{itemize}

We use $a_x^{\prime}$ to highlight that a general payment stream is assumed; i.e. the payments of amount 1 are made at some fraction $f \in \left[ 0,1 \right]$ through the year, and hence $a_x$ and $\ddot{a}_x$ are simply special cases of $a_x^{\prime}$ with $f=1$ and $f=0$ respectively.

In a similar manner, standard notation uses $a_{x|y}$ and $\ddot{a}_{x|y}$ to represent two-life reversionary annuities where the payments are made at the end or start of a year, respectively, to $(y)$ after the death of $(x)$ (where $(z)$ denotes a life aged exactly $z$).  Here, we use $a_{x|y}^{\prime}$ to indicate that payments are made some fraction $f\in[0,1]$ through the year so that $a_{x|y}$ and $\ddot{a}_{x|y}$ are just special cases of $a_{x|y}^{\prime}$.

Finally, we use $v$ to denote the discount factor which applies over a period of unit length, so that $v^{t+f}$ is the discount factor which applies from the point of valuation to a cash flow some fraction $f \in \left[ 0,1 \right]$ through the step from $t$ to $t+1$ for $t \in \mathbb{Z}^{+}$.

\subsection{Annuities}
     \label{Annuities}

At its most basic, an annuity is just a stream of payments, and a life annuity is an annuity where the payments depend on the survival, or otherwise, of a pre-specified life, or lives.  Section \ref{Single-LifeAnnuityCertain} considers single-life annuities certain; they depend on the survival, or death, of only one life, and they will start to be paid.  Other types of life annuity could depend on two or more lives, and yet others may not even be start to be paid; for example, a child's annuity which is a rider benefit to a temporary assurance of the parent will only come into payment if that parent dies within the period specified in the contract.  Section \ref{Two-LifeReversionaryAnnuity} considers two-life reversionary annuities since this is the contract which provided our motivation for developing a vector form of a recurrence relation.

\subsubsection{Single-Life Annuity Certain}
     \label{Single-LifeAnnuityCertain}

Annuities may be paid in arrears, or in advance, although many annuities are paid part way through each period; obvious examples are pensions where the payments are often made on the `{\it monthiversary}' of the policy inception date.  In Section \ref{ActuarialNotation} it was noted that the advance and arrear cases are just special cases of annuities where payments are made part way through each period; it is therefore appropriate to we only consider cases where payments are made at a fraction $f \in [0,1]$ through the step.

Level, single-life annuities serve as an introduction to several concepts which become useful later, particularly when considering other types of annuity.  For a general, single-life, level annuity the summation formula for the reserve factor is
\begin{equation}
    a^{\prime}_x = \sum_{t=0}^{\infty} {}_{t+f}p_x \: \times \: v^{t+f} \quad \text{for } f \in \left[0,1 \right]      \label{eq:slAnnSumm}
\end{equation}

The derivation of the recurrence relation is straightforward:
\begin{align*}
     a^{\prime}_x & = \sum_{t=0}^{\infty} {}_{t+f}p_x \: \times \: v^{t+f}
  \\         & = {}_{f}p_x \: \times \: v^{f} + \sum_{(s+1)=1}^{\infty} {}_{\left((s+1)+f\right)}p_x \: \times \: v^{\left((s+1)+f\right)}
  \\         & = {}_{f}p_x \: v^{f} + p_x \: v \: \sum_{s=0}^{\infty} {}_{\left(s+f\right)}p_{x+1} \: \times \: v^{\left(s+f\right)}
\end{align*}
and, by comparing the summation with Equation \ref{eq:slAnnSumm}, the recurrence relationship is
\begin{equation}
    \boxed{
            a^{\prime}_x = {}_{f}p_x \: v^{f} + p_x \: v \: a^{\prime}_{x+1}
          }
    \label{eq:slAnnRecRel}
\end{equation}
Note that this derivation assumes that both mortality rates and interest rates are constant in time; in Section \ref{Assumptions} these assumptions will be removed.

For an annuity payable annually in advance, this relation reduces to \[ \ddot{a}_{x} = 1 + v \: p_{x} \: \ddot{a}_{x+1} \]

Note that, when written in the form of Equation \ref{eq:slAnnRecRel}, the recurrence runs backwards in time; this is convenient since natural boundary conditions exist (or may be assumed) at the end of the policy, e.g. $p_{120}=0$.

A useful side-effect of using the recurrence is that it removes the need to use the power function to compute $v^{t+f}$ in Equation \ref{eq:slAnnSumm}, and replaces it with multiplication, which is about 20 times cheaper in modern hardware.

\subsubsection{Two-Life Reversionary Annuity}
     \label{Two-LifeReversionaryAnnuity}
Re-interpreting a standard definition \citep[Section 8.6]{Neill}, a {\it reversionary annuity} ``becomes payable on the failure of a specified status and remains payable during the continued existence of a second status''.  The simplest form of reversionary annuity is one which becomes payable to $(y)$ on the death of $(x)$, if $(y)$ is alive at that time, and remains payable until the death of $(y)$; a common example is a spouse's pension which becomes payable to a pensioner's spouse on the death of that pensioner.

Suppose that the lives are aged $x$ and $y$, and that they are independent.  Assume, without loss of generality, that the payment is made to $(y)$ after the death of $(x)$. Then
\begin{align*}
    \text{Pr[payment at time $t$ is made]}
        & = \text{Pr[$x$ is dead at time $t$ and $y$ is alive at time $t$]}
 \\     & = \text{Pr[$x$ dies within $t$ years] $\times$ Pr[$y$ survives for $t$ years]}
 \\     & = (1-{}_tp_x) \times {}_tp_y
\end{align*}

For the case where level payments are made part-way through an interval, at some fraction $f$ from the start of each interval, the summation formula for the reserve factor is
\begin{equation}
    a^{\prime}_{x|y}
        = \sum_{t=0}^{\infty} \left(1-{}_{t+f}p_{x}\right) \times {}_{t+f}p_{y} \times v^{t+f} \quad \text{for } f \in \left[ 0,1 \right]
        \label{eq:revAnnSumm}
\end{equation}

Again, the recurrence relation may be derived straightforwardly:
\begin{align*}
    a^{\prime}_{x|y}
        &= \sum_{t=0}^{\infty} \: \left( 1- {}_{t+f}p_{x} \right) \cdot {}_{t+f}p_{y} \cdot v^{t+f}
  \\    &= {}_{f}q_{x} \cdot {}_{f}p_{y} \cdot v^{f}
         \quad
         + \sum_{(s+1)=1}^{\infty} \: {}_{(s+1)+f}p_{y} \cdot v^{(s+1)+f}
  \\    &\quad \quad \quad \quad \quad \quad \quad \quad
         - \sum_{(s+1)=1}^{\infty} \: {}_{(s+1)+f}p_{x} \cdot {}_{(s+1)+f}p_{y} \cdot v^{(s+1)+f}
  \\    &= {}_{f}q_{x} \cdot {}_{f}p_{y} \cdot v^{f}
         \quad
         + v \: p_{y} \cdot a^{\prime}_{y+1}
  \\    &\quad \quad \quad \quad \quad \quad \quad \quad
         - v \: p_{x} \: p_{y} \: \sum_{s=0}^{\infty} \: \left[ 1-{}_{s+f}q_{x+1} \right] \cdot
                                       {}_{s+f}p_{y+1} \cdot v^{s+f}
\end{align*}
i.e.
\begin{equation}
    \boxed{
       a^{\prime}_{x|y} = {}_{f}q_{x} \: {}_{f}p_{y} \: v^{f}
                        + {}_{}q_{x} \: {}_{}p_{y} \: v \: a^{\prime}_{y+1}
                        + {}_{}p_{x} \: {}_{}p_{y} \: v \: a^{\prime}_{x+1|y+1}
          }
    \label{eq:revAnnRecRel}
\end{equation}

In turn, the three components represent
\begin{enumerate}[1)]
  \item the payment at time $f$ which is made only if $x$ has died but $y$ is still alive,
  \item the (single life) reserve factor at the start of the next step if $x$ has died but $y$ is still alive, and
  \item the (reversionary) reserve factor at the start of the next step if both $x$ and $y$ are still alive.
\end{enumerate}

\subsection{Vector Recurrence Relation}
     \label{VectorRecurrence}
The derivation of the vector form of the recurrence relation is based on the observation that the recurrence relation for the reserve factor for the reversionary annuity in Equation \ref{eq:revAnnRecRel} also involves the reserve factor for the single life annuity.

\subsubsection{Contract-Based Presentation}
        \label{ContractBasedPresentation}
When expressed as a vector, the pair of recurrence relations in Equations \ref{eq:revAnnRecRel} and \ref{eq:slAnnRecRel} become
\begin{align*}
         \begin{pmatrix}
              a^{\prime}_{x|y}
                       \\ ~ \\
              a^{\prime}_{y}
         \end{pmatrix}
      &= \begin{pmatrix}
              {}_{f}q_x \: {}_{f}p_y \: v^{f}
                    + p_y \: q_x \: v \: a^{\prime}_{y+1}
                    + p_x \: p_y \: v \: a^{\prime}_{x+1|y+1}
                       \\ ~ \\
              {}_{f}p_y \: v^{f}
                    + p_y \: v \: a^{\prime}_{y+1}
         \end{pmatrix}
\end{align*}
which can be written as
\begin{equation}
       \begin{pmatrix}
            a^{\prime}_{x|y}
                     \\ ~ \\
            a^{\prime}_{y}
       \end{pmatrix}
    = v^{f} \:
       \begin{pmatrix}
            {}_{f}q_x \: {}_{f}p_y  &  0    \\ ~ \\    0  &  {}_{f}p_y
       \end{pmatrix}
       \begin{pmatrix}
            1   \\ ~ \\   1
       \end{pmatrix}
     + v \:
       \begin{pmatrix}
          p_x \: p_y  &  p_y \: q_x
                     \\ ~ \\
            0  &  p_y
       \end{pmatrix}
       \begin{pmatrix}
            a^{\prime}_{x+1|y+1}
                     \\ ~ \\
            a^{\prime}_{y+1}
       \end{pmatrix}
       \label{eq:revAnnRecRelVec}
\end{equation}

Hence, for a relatively simple contract, it is possible to find a vector expression for the recurrence relation, where that vector expression brings together the relations for all contract types which may be involved in the reserve calculations for that policy type.

\subsubsection{Survival-Based Presentation}
        \label{SurvivalBasedPresentation}

Although the two matrices in Equation \ref{eq:revAnnRecRelVec} are the same shape, they are populated differently; this results from the derivation being based on the type of the policy at each step.  An alternative presentation is to consider the survival state of each life at each step.  Using binary indexing for each life being either alive (state 0) or dead (state 1) leads to a simple representation of all the possibilities of survival over the step; the states and labelling for two lives, currently aged $x_1$ and $x_2$, are therefore
  \begin{center}
    \begin{tabular}{c c c c}
      State  &  Label  &  $(x_1)$  &  $(x_2)$    \\
      \hline
      0  &  $S_{00}$  &  Alive  &  Alive  \\
      1  &  $S_{01}$  &  Alive  &  Dead  \\
      2  &  $S_{10}$  &  Dead  &  Alive  \\
      3  &  $S_{11}$  &  Dead  &  Dead  \\
    \end{tabular}
  \end{center}

Using the ordering which results from this binary labelling, it is possible to construct a matrix of probabilities of the lives surviving for time $g$, which could be either $f$ (when considering the probability of payment), or $1$ (when considering the probability of requiring a reserve).  For two lives, the relevant Markov transition matrix is
\begin{equation*}
    \begin{pmatrix}
      {}_{g}p_{x} \: {}_{g}p_{y}  &  {}_{g}p_{x} \: {}_{g}q_{y}  &  {}_{g}q_{x} \: {}_{g}p_{y}  &  {}_{g}q_{x} \: {}_{g}q_{y}
      \\  ~  \\
      0  &  {}_{g}p_{x}   &  0  &  {}_{g}q_{x}
      \\  ~  \\
      0  &  0  &  {}_{g}p_{y}  &  {}_{g}q_{y}
      \\  ~  \\
      0  &  0  &  0  &  1
    \end{pmatrix}
\end{equation*}
The states of the lives can be considered as a time-inhomogeneous Markov chain, where this is the transition matrix if $g=1$.

Using this matrix to combine Equations \ref{eq:revAnnRecRel} and \ref{eq:slAnnRecRel}, and setting $g=f$ or $g=1$, leads to the following recurrence for a reversionary annuity
\begin{align*}
       \begin{pmatrix}
            a^{\prime}_{x|y}   \\ ~ \\   0   \\ ~ \\   a^{\prime}_{y}   \\ ~ \\   0
       \end{pmatrix}
    = v^{f} \:
       \begin{pmatrix}
            {}_{f}p_{x} \: {}_{f}p_{y}  &  {}_{f}p_{x} \: {}_{f}q_{y}  &  {}_{f}q_{x} \: {}_{f}p_{y}  &  {}_{f}q_{x} \: {}_{f}q_{y}
            \\  ~  \\
            0  &  {}_{f}p_{x}   &  0  &  {}_{f}q_{x}
            \\  ~  \\
            0  &  0  &  {}_{f}p_{y}  &  {}_{f}q_{y}
            \\  ~  \\
            0  &  0  &  0  &  1
       \end{pmatrix}
       \begin{pmatrix}
            0   \\ ~ \\   0   \\ ~ \\   1   \\ ~ \\   0
       \end{pmatrix}
     + v \:
       \begin{pmatrix}
            p_{x} \: p_{y}  &  p_{x} \: q_{y}  &  q_{x} \: p_{y}  &  q_{x} \: q_{y}
            \\  ~  \\
            0  &  p_{x}  &  0  &  q_{x}
            \\  ~  \\
            0  &  0  &  p_{y}  &  q_{y}
            \\  ~  \\
            0  &  0  &  0  &  1
       \end{pmatrix}
       \begin{pmatrix}
            a^{\prime}_{x+1|y+1}   \\ ~ \\   0   \\ ~ \\   a^{\prime}_{y+1}   \\ ~ \\   0
       \end{pmatrix}
\end{align*}
This vector expression for the reserve factors under consideration appears more complex than the version in Equation \ref{eq:revAnnRecRelVec}; however, this expression involves only one matrix, which must be evaluated at two points in time, rather than requiring two different matrices, one for each of the time points.

Therefore, although binary labelling using `zero for alive' may initially seem counter-intuitive, its use leads to a desirable property; since the number of dead people cannot decrease, this labelling will naturally lead to the transition matrix being upper triangular.

\subsection{Summary}
     \label{SummaryDerivation}
This Section has provided our motivation; it has demonstrated that, for simple cases, it is possible to produce a vector recurrence relation for successive reserve factors, using two instances of the same matrix, where that matrix is expressed in terms of transitions between survival states of the lives involved, rather than the contract types involved.

%% file: derivation.tex
\section{Derivation of Vector Relation}
  \label{Derivation}

In this section we present a completely general vector recurrence relation which may be used for different forms of insurance contract; after the initial presentation, we formally derive the relation using a component-wise approach.  The generic nature of the components in the recurrence relation means that this vector relation can be used for any policy of arbitrary complexity, so long as the cash flows and probabilities can be isolated and expressed in the form required by the vector notation presented here.

\subsection{General Form}
     \label{GeneralForm}
The relations expressed in Equation \ref{eq:revAnnRecRelVec} are a specific case of the general form which may be stated as
\begin{equation}
   \boxed{
           \mathbf{r}_{\mathbf{x},t}
                  = v_{t}^f \: \mathbf{W}_{\mathbf{x},t,f} \: \mathbf{c}_{\mathbf{x},t}
                  + v_{t}   \: \mathbf{W}_{\mathbf{x},t,1} \: \mathbf{r}_{\mathbf{x+1},t}
           \quad \quad
           \forall \: t \in \mathbb{R}^{+}
           \quad
           f \in [0,1]
         }
   \label{eq:generalForm}
\end{equation}
where
\begin{itemize}
  \item[-]  $\mathbf{x+t}$ is a vector representing the ages, at time $t$, of a collection of $m$ lives,

  \item[-]  $n$ is the number of states of survivorship which need to be considered for the $m$ lives in $\mathbf{x}$; all the states must be considered in the formation of the complete vector recurrence relation,

  \item[-]  $\mathbf{r}_{\mathbf{x},t}$ is a vector of $n$ reserve factors, at time $t \ge 0$, which may be required depending on the survival state of $\mathbf{x}$ at that time,
    \\   n.b. since the relation involves $\mathbf{r}_{\mathbf{x},t}$ and $\mathbf{r}_{\mathbf{x+1},t}$, the reserves are calculated using steps of unit interval,

  \item[-]  $f\in [0,1]$ is the fraction through a step at which any cash flows occur,

  \item[-]  $g \in \{f,1\}$ is the fraction through the step at which events of importance occur; any cash flows which occur happen at fraction $f$, and reserve factors relating to the end of a step of unit length are required when $g=1$,

  \item[-]  $v_{t}^g$, where $g \in \{f,1\}$, is the discount factor which applies from time $t$ either to any cash flows which may be made (at time $t+f$), or to the reserve factors relating to the end of a step of unit length (so that $t+1$); the underlying interest rate may vary with time,

  \item[-]  $\mathbf{W}_{\mathbf{x},t,g}$ is an $n \times n$ Markov transition matrix containing the probabilities of the lives surviving from time $t$ up to time $t+g$; the underlying mortality may vary with time,

  \item[-]  $\mathbf{c}_{\mathbf{x},t}$ is a column vector of length $n$ containing the nominal amounts of the cash flows which may be made at time $t+f$, depending on the state of the lives at that time,

  \item[-]  units of time are arbitrary; in practice units are often months or years.
\end{itemize}
Since, in Equation \ref{eq:generalForm}, $\mathbf{r}_{\mathbf{x},t}$ is expressed in terms of $\mathbf{r}_{\mathbf{x+1},t}$, numerical values for each element of each vector may be obtained by backward substitution, starting from an appropriate boundary condition.

Using a standard definition \citep[Section 4.2]{Neill},
\begin{quote}
  A (prospective) reserve is the present value of all future cash flows, allowing for discounting and the probability of those cash flows being made.
\end{quote}
The vector approach presented here is confirmation of the intuitive interpretation of that definition, i.e.
\begin{quote}
  {\it The reserve `now' is the present value of `any cash flows which may occur during the first period' together with the present value of `any reserve which is required at the end of the first period, so long as that reserve is then required'}.
\end{quote}

\subsection{The Lives}
     \label{TheLives}

In theoretical work, it is standard practice to assume that all lives involved in a policy are independent.  For the most general case, there could be any number of lives, so we consider $\mathbf{x+t}$, which represents a collection of $m$ independent lives aged precisely $x_1+t,x_2+t,\ldots{},x_m+t$; this notation reinforces the fact that the ages of the lives are a function of the time $t$ since the start of the projection, where $\mathbf{t}$ is a vector, of length $m$, whose elements are all equal to $t$.  Using this notation, the ages of a given set of lives are just a function of the duration from valuation to the time of interest.

\subsection{Cash Flows}
     \label{CashFlows}
We are interested in a sequence of cash flows which may be made at fixed future times, according to the survival state of $\mathbf{x}$ at those future times.  The nominal amount of each cash flow is fixed, but the expected amount of each cash flow depends on the probability that the payment is made and hence on the survival state, at the point of payment, of the collection of lives.

Let $t \in \mathbb{R}^{+}$ and let $\beta \in \mathbb{R}^{+}$.  Let $c_{\mathbf{x},t,j}$ be the nominal cash flow which happens at the fraction $f \in [0,1]$ from time $t$, if $\mathbf{x}$ is in state $j$ at that time.  Then
\begin{align*}
                 & \text{ the cash flow $c_{\mathbf{x},t,j}$ happens at time $t+f$ if the lives are in state $j$ at that time}
  \\ \Rightarrow & \text{ the cash flow $c_{\mathbf{x},t,j}$ happens at time $(t-\beta)+\beta+f$ if the lives are in state $j$ at that time}
  \\ \Rightarrow & \text{ the cash flow $c_{\mathbf{x},t,j}$ may also be denoted $c_{\mathbf{x+\boldsymbol\beta},t-\beta,j}$
}
\end{align*}
i.e.
\[
      c_{\mathbf{x},t,j} = c_{\mathbf{x+\boldsymbol\beta},t-\beta,j}
      \quad \quad
      \forall \: \mathbf{x}
      \quad \quad
      0 < \beta < t
\]
This is the mathematical formulation of the statement that ``a payment of known nominal amount, made at some time in the future, will be of the same nominal amount at that future date, irrespective of how the time in the future is determined''.

Therefore, in particular, let $\beta=1$ and let $t=s+1$.  Then
\begin{equation}
  \boxed{
          c_{\mathbf{x},s+1,j} = c_{\mathbf{x+1},s,j}
        }
  \label{recRel_CashFlows}
\end{equation}
It should be noted that the cash flows are a completely general, arbitrary function of the state of the lives at the time the payment is made; also see Section \ref{Assumptions_CashFlows}.

\subsection{Discount Factors}
     \label{DiscountFactors}
In standard notation, if $i$ is a constant annual rate of interest which applies over a period of $t$ years then the discount factor which applies for that period is $v^t = (1+i)^{-t}$.  There is an implicit assumption that the period starts at time $0$ and ends at at time $t$; if that is not the case then the discount factor remains $v^t = (1+i)^{-t}$ wherever the difference in time is $t$.  Hence, when the rate of interest is constant, $v^t = v^{t_1} \: v^{t-t_1}$ for any $0 \le t_1 \le t$.

When the rate of interest varies with time, standard practice \citep[Section 2.4]{McCutcheon_Scott} is to consider the discount factor from $0$ to $t$ as
\[
    v(t) = \exp \left(-\int_{0}^{t} \delta(r) \cdot dr \right)
\]
where $\delta(r)$ is the {\it force of interest} at time $r$.

Let $0 \le t_1 \le t_2$, and let $\delta(r)$ be the force of interest at time $r \in [0,t_2]$.
Let $d_{\mathbf{x},t,t_1,t_2}$ be the discount factor which applies from time $t+t_1$, and lasts for time $t_2-t_1$; i.e. it applies from the point at which the lives are aged $\mathbf{x+t+t_1}$ to the point the where lives are aged $\mathbf{x+t+t_2}$.
Then
\begin{align*}
  d_{\mathbf{x},t,t_1,t_2}
    &= \exp \left(-\int_{t_1}^{t_2} \delta(r) \cdot dr \right)
  \\  &= \exp \left(- \left[ \int_{t_1}^{t^{\prime}} \delta(r) \: dr + \int_{t^{\prime}}^{t_2} \delta(r) \: dr \right] \right)
  \\  &= \exp \left(- \int_{t_1}^{t^{\prime}} \delta(r) \: dr \right)  \times  \exp \left(- \int_{t^{\prime}}^{t_2} \delta(r) \: dr \right)
\end{align*}
i.e.
\[
      d_{\mathbf{x},t,t_1,t_2} = d_{\mathbf{x},t,t_1,t^{\prime}} \cdot d_{\mathbf{x},t,t^{\prime},t_2}
      \quad \quad
      \forall \: \mathbf{x}
      \quad \quad
      t_1 \le t^{\prime} \le t_2
\]
This is the mathematical formulation of the statement that ``discounting an amount from $t_2$ to $t_1$ is the same as discounting that amount from $t_2$ to an intermediate time $t^{\prime}$ and then discounting that (reduced) amount from $t^{\prime}$ to $t_1$''.

Therefore, in particular, let $t_1=0$, $t^{\prime}=1$ and $t_2=s+1+f$ so that $t_2-t_1=s+1+f$, $t^{\prime}-t_1=1$ and $t_2-t^{\prime}=s+f$.  Then
\begin{equation}
  \boxed{
          d_{\mathbf{x},t,0,s+1+f} = d_{\mathbf{x},t,0,1} \cdot d_{\mathbf{x},t,1,s+f}
        }
  \label{recRel_Discounts}
\end{equation}
Notice that these discount factors are independent of the state of the lives, a phenomenon which is consistent with reality since mortality and investment returns are generally independent.  Notice also that the use of force of interest removes any assumption that the interest rate remains constant; also see Section \ref{Assumptions_Interest}.

\subsection{Survival Probabilities}
     \label{SurvivalProbabilities}

In standard notation, ${}_{t}p_{x}$ is the probability that a life aged precisely $x$ survives for $t$ years.  Hence, the probability that two independent lives aged precisely $x$ and $y$ both survive for $t$ years is denoted ${}_{t}p_{x} \cdot {}_{t}p_{y}$.  By extension, the probability that a collection of $m$ independent lives all survive for $t$ years is simply the product of the survival probabilities of each of the individual lives.

It is necessary to consider the probability of moving from any state to any other.  Let $w_{\mathbf{x},t,t_1,t_2,j,i}$ be the probability of the set lives $\mathbf{x}$ being in state $i$ at time $t+t_2$, given that it is in state $j$ at time $t+t_1$ for $t_1 \le t_2$ (so that the time for the possible transition between states is $t_2-t_1$).  
Then, using the Partition Theorem, and conditioning on the state at time $t^{\prime} \in [0,t_2-t_1]$,
\[
      w_{\mathbf{x},t,t_1,t_2,j,i} = \sum_k w_{\mathbf{x},t,t_1,t^{\prime},j,k} \cdot w_{\mathbf{x+t^{\prime}},t^{\prime},t^{\prime},t_2,k,i}
\]
where the sum is over all possible states to which the lives could migrate.
This is the mathematical formulation of the statement that ``the probability of the set of lives moves from state $j$ to state $i$ in time $t$ is the same as the probability of moving from state $j$ to any other state at time $t^{\prime}$ and then moving into state $i$ in the remaining time''.

Therefore, in particular, let $t_1=0$, $t^{\prime}=1$ and $t_2=s+1+f$ so that $t_2-t_1=s+1+f$, $t^{\prime}-t_1=1$ and $t_2-t^{\prime}=s+f$.  Then
\begin{equation}
  \boxed{
      w_{\mathbf{x},t,0,s+1+f,j,i} = \sum_k w_{\mathbf{x},t,0,1,j,k} \cdot w_{\mathbf{x+1},t,1,s+f,k,i}
        }
  \label{recRel_Probabilities}
\end{equation}
Notice that there is no assumption of constant mortality; also see Section \ref{Assumptions_Mortality}.

\subsection{Recurrence Relation for Reserve Factors}
\label{Recurrence Relation}

Using the results in the preceding Sections it is possible to derive a general recurrence relation where benefits may be payable, depending on the survival state of the lives.  This is a two stage process; first, a relation is derived for lives being in a particular state, and then the general relation is obtained by considering all possible states that the lives may be in.

\subsubsection{Reserve for Lives in a Particular State}
        \label{Reserve_CurrentState}

From Section \ref{CashFlows}, any cash flows which happen in a step happen at some fraction $f \in [0,1]$ through the step.  Let $\mathbf{r}_{\mathbf{x},t,j}$ be the reserve which must be held, at time $t$, for a set of lives $\mathbf{x}$ in state $j$ at that time.  Then, because the lives could have migrated to state $i$ by time $s$, the nominal amount of the cash flow to be made in step $t+s$ to the lives in $\mathbf{x}$, if they are in state $i$ at that time is $c_{\mathbf{x},t+s,i}$.  Therefore, the expected amount of a cash flow to be made in step $t+s$ to the lives in $\mathbf{x}$, given that they are currently in state $j$, and allowing for migration to any other state $i$, is
\[
    \sum_{i}
    \left[
        w_{\mathbf{x},t,0,s+f,j,i}
        \cdot{}
        c_{\mathbf{x},t+s,i}
    \right]
\]
Allowing for discounting to that time, the present value of such a cash flow is
\[
    d_{\mathbf{x},t,0,s+f}
    \sum_{i}
    \left[
        w_{\mathbf{x},t,0,s+f,j,i}
        \cdot{}
        c_{\mathbf{x},t+s,i}
    \right]
\]
Therefore, using an infinite sum to allow for possible cash flows in all future time steps, the reserve required at time $t$, given that the lives $\mathbf{x}$ are in state $j$ at that time, is
\begin{equation}
  r_{\mathbf{x},t,j}
      = \sum_{s=0}^{\infty}
          \left(
              d_{\mathbf{x},t,0,s+f}
              \sum_{i}
              \left[
                  w_{\mathbf{x},t,0,s+f,j,i}
                  \cdot{}
                  c_{\mathbf{x},t+s,i}
              \right]
          \right)
    \label{basicReserveSummation}
\end{equation}
for $t=0,1,2,\cdots,\infty$.  For the brute force approach, the 1000 simulations at each time step $t$ will each be an instance of $r_{\mathbf{x},t,j}$.

\subsubsection{Relation for Lives in a Particular State}
\label{Relation_CurrentState}
Starting from Equation \ref{basicReserveSummation}, the recurrence relation for lives in a particular state may be derived in a straightforward manner;
\begin{align}
    r_{\mathbf{x},t,j}
      & = \sum_{s=0}^{\infty}
          \left(
              d_{\mathbf{x},t,0,s+f}
              \sum_{i}
              \left[
                  w_{\mathbf{x},t,0,s+f,j,i}
                  \cdot{}
                  c_{\mathbf{x},t+s,i}
              \right]
          \right)
        \label{reserve_summation}
\intertext{which, splitting off the first term in the sum, is}
      & = d_{\mathbf{x},t,0,f}
          \sum_{i}
          w_{\mathbf{x},t,0,f,j,i}
          \cdot{}
          c_{\mathbf{x},t,i}
        \nonumber
  \\  &   \quad \quad \quad
        + \sum_{s=1}^{\infty}
          \left(
              d_{\mathbf{x},t,0,s+f}
              \sum_{i}
              w_{\mathbf{x},t,0,s+f,j,i}
              \cdot{}
              c_{\mathbf{x},t+s,i}
          \right)
        \nonumber
\intertext{and, shifting the index (so that $s = s^{\prime}+1$), this is}
      & = d_{\mathbf{x},t,0,f}
          \sum_{i}
          w_{\mathbf{x},t,0,f,j,i}
          \cdot{}
          c_{\mathbf{x},t,i}
        \nonumber
  \\  &   \quad \quad \quad
        + \sum_{s^{\prime}=0}^{\infty}
          \left(
              d_{\mathbf{x},t,0,s^{\prime}+1+f}
              \sum_{i}
              w_{\mathbf{x},t,0,s^{\prime}+1+f,j,i}
              \cdot{}
              c_{\mathbf{x},t+s^{\prime}+1,i}
          \right)
        \nonumber
\intertext{which, using the results from Equations \ref{recRel_CashFlows}, \ref{recRel_Discounts}, and \ref{recRel_Probabilities}, is}
      & = d_{\mathbf{x},t,0,f}
          \sum_{i}
          w_{\mathbf{x},t,0,f,j,i}
          \cdot{}
          c_{\mathbf{x},t,i}
        \nonumber
  \\  &   \quad \quad \quad
        + \sum_{s^{\prime}=0}^{\infty}
          \left(
              \left[
                d_{\mathbf{x},t,0,1} \cdot d_{\mathbf{x},t,1,s^{\prime}+f}
              \right]
              \sum_{i}
              \left[
                \sum_k w_{\mathbf{x},t,0,1,j,k} \cdot w_{\mathbf{x+1},t,1,s^{\prime}+f,k,i}
              \right]
              \cdot{}
              \left[
                c_{\mathbf{x+1},t+s^{\prime},i}
              \right]
              \right)
        \nonumber
\intertext{and, reordering the summations and factoring terms which are sum independent, this is}
      & = d_{\mathbf{x},t,0,f}
          \sum_{i}
          w_{\mathbf{x},t,0,f,j,i}
          \cdot{}
          c_{\mathbf{x},t,i}
      \nonumber
  \\  &  \quad \quad \quad
        + d_{\mathbf{x},t,0,1} \:
          \sum_k
          w_{\mathbf{x},t,0,1,j,k}
          \left[
             \sum_{s^{\prime}=0}^{\infty}
             d_{\mathbf{x},t,1,s^{\prime}+f}
             \sum_{i}
             \left(
                 w_{\mathbf{x+1},t,1,s^{\prime}+f,k,i}
                 \cdot{}
                 c_{\mathbf{x+1},t+s^{\prime},i}
             \right)
          \right]
      \nonumber
\end{align}
Finally, recognising that $d_{\mathbf{x},t,1,s^{\prime}+f} = d_{\mathbf{x+1},t,0,s^{\prime}+f}$, and comparing sum in square brackets to the summation required for $\mathbf{r}_{\mathbf{x+1},t,k}$, gives
\begin{equation}
    r_{\mathbf{x},t,j}
      =
        d_{\mathbf{x},t,0,f}
        \sum_{i}
        w_{\mathbf{x},t,0,f,j,i}
        \cdot{}
        c_{\mathbf{x},t,i}
      +
        d_{\mathbf{x},t,0,1} \:
        \sum_k
        w_{\mathbf{x},t,0,1,j,k}
        \cdot{}
        r_{\mathbf{x+1},t,k}
        \label{eq:generalRecurrence}
\end{equation}
where the $i$ and $k$ sums are over all possible states of the lives.

Various policy types could fit a particular instance of this formula just by changing the values of $c_{\mathbf{x},t,i}$ which may or may not be zero, depending on the state $j$.  Notice that a policy which has a limited term trivially fits the use of an infinite sum by setting the cash flow amounts after the end of the policy term to zero.

\subsubsection{Relation Considering all Possible States}
\label{Relation_AllStates}

Equation \ref{eq:generalRecurrence} relates to the $j$\textsuperscript{th} possibility of a set of possible states. Combining all of the possibilities for $r_{\mathbf{x},t,j}$ into a vector, the relationship becomes
\begin{equation*}
    \mathbf{r}_{\mathbf{x},t}
        =  d_{\mathbf{x},t,0,f} \: \mathbf{W}_{\mathbf{x},t,f} \: \mathbf{c}_{\mathbf{x},t}
        +  d_{\mathbf{x},t,0,1} \: \mathbf{W}_{\mathbf{x},t,1} \: \mathbf{r}_{\mathbf{x+1},t}
\end{equation*}
where
\begin{itemize}
  \item[-]  $\mathbf{r}_{\mathbf{x},t}$ is a column vector, of length $n$, where the $i$\textsuperscript{th} entry is $r_{\mathbf{x},t,i}$,

  \item[-]  $d_{\mathbf{x},t,0,g}$ is the discount factor from time $t$ up to time $t+g$ for $g \in \left\{ f,1 \right\}$,

  \item[-]  $\mathbf{W}_{\mathbf{x},t,g}$ is an $n \times n$ Markov transition matrix where the entries relate to the lives surviving from time $t$ to the time $t+g$ for $g \in \left\{ f,1 \right\}$; the $(i,j)$\textsuperscript{th} entry of $\mathbf{W}_{\mathbf{x},t,g}$ is $w_{\mathbf{x},t,0,g,j,i}$,

  \item[-]  $\mathbf{c}_{\mathbf{x},t}$ is a column vector, of length $n$, where the $i$\textsuperscript{th} entry is $c_{\mathbf{x},t,i}$.
\end{itemize}
Replacing the discount factors $d_{\mathbf{x},t,0,g}$ with $v_{t}^g$ which is closer to the equivalent standard notation, recognises that interest rates do not depend on the state of any lives, and explicitly allows the possibility that the underlying interest rate varies through the projection, the equation becomes
\begin{equation*}
           \mathbf{r}_{\mathbf{x},t}
                  = v_{t}^f \: \mathbf{W}_{\mathbf{x},t,f} \: \mathbf{c}_{\mathbf{x},t}
                  + v_{t}   \: \mathbf{W}_{\mathbf{x},t,1} \: \mathbf{r}_{\mathbf{x+1},t}
           \quad \quad
           \forall \: \mathbf{x} \in \mathbb{R}^{m^{+}}
           \quad \quad
           f \in [0,1]
\end{equation*}
as presented in Equation \ref{eq:generalForm}.

Death is an absorbing state, so that if the lives are in state $k$ at time $0$ and still in state $k$ at time $t>0$ then they must have been in that state at all intermediate times.  Also, if there are two or more changes of state in a step, then we  are only interested in the overall change; e.g. if a reversionary annuity changes to a single life annuity on the death of the first life, and becomes ceased on the death of the second life within the same time step, then we are only interested in the fact that it has gone from being a reversionary annuity to being ceased, and the fact that it was temporarily a single life annuity is of no consequence.

\subsubsection{Zero Reserve States}
     \label{Zero Reserve States}

We call a state a `zero reserve state' (ZRS) if it is a state for which all future cash flows are zero and there is no path to a state which has any non-zero cash flows.  There is no need to keep track of ZRSs (since they do not contribute to the liabilities) and hence calculations which relate to ZRSs are redundant and may be removed.
For example, a reversionary annuity has a zero cash flow if $(x)$ is still alive, but the state where $(x)$ and $(y)$ are both alive is not a ZRS because there is a path to a state where there are future cash flows, i.e. the path to the state where $(x)$ dies before $(y)$.

A ZRS is not necessarily a sink state for the Markov chain because it is possible to move out of one ZRS into another ZRS; for example, a reversionary annuity has no future cash flows when $(y)$ dies so that the state where $(y)$ dies first is a ZRS from which it is possible to move into another ZRS (on the death of $(x)$).  Allowing for ZRSs, Equation \ref{eq:generalForm} may be written as
\begin{equation*}
           \overline{\mathbf{r}}_{\mathbf{x},t}
                  = v_{t}^f \: \overline{\mathbf{W}}_{\mathbf{x},t,f} \: \overline{\mathbf{c}}_{\mathbf{x},t}
                  +  v_{t}  \: \overline{\mathbf{W}}_{\mathbf{x},t,1} \: \overline{\mathbf{r}}_{\mathbf{x+1},t}
\end{equation*}
where
\begin{itemize}
  \item[-]  $\overline{\mathbf{r}}_{\mathbf{x},t}$ is $\mathbf{r}_{\mathbf{x},t}$ with ZRSs removed,

  \item[-]  $\overline{\mathbf{c}}_{\mathbf{x},t}$ is $\mathbf{c}_{\mathbf{x},t}$ with ZRSs removed, and

  \item[-]  $\overline{\mathbf{W}}_{\mathbf{x},t,g}$ is $\mathbf{W}_{\mathbf{x},t,g}$ with rows and columns which correspond to ZRSs removed.
\end{itemize}

\subsection{Assumptions}
     \label{Assumptions}
The derivations in the preceding sections require only the simplest of assumptions which, for practical purposes, are not particularly restrictive.

\subsubsection{Cash Flows}
     \label{Assumptions_CashFlows}
The monetary amounts of all cash flows (whether they are premiums, benefits or expenses) must be known in advance of their use.  Also, more as a requirement to be able to use vector arithmetic than an assumption, there must be a one-to-one correspondence between time steps and cash flows (so that there can be a maximum of one cash flow of any particular type in each projection step).  Therefore, if a policy has monthly cash flows then monthly projection steps are required; using monthly projection steps is fine for policies which have annual cash flows since 11 of any 12 consecutive steps will have a cash flow of amount zero.

\subsubsection{Interest}
     \label{Assumptions_Interest}
The interest rate being used for a particular step must be known in advance of reaching the time step being simulated.  Since the derivation in Section \ref{DiscountFactors} is based on the force of interest, there is no need to assume that the interest rate is constant.  There is, however, a requirement that the time interval under consideration can be split appropriately and, for all practical purposes, this should be possible.  Treating inflation as `the other side of the interest coin' requires that an equivalent assumption applies to inflation of expenses.

There is a great body of research into modelling future interest rates; see for example \cite{Vasicek_Interest}, \cite{CoxIngersoll_Interest} and \cite{Chen_Interest}. Any interest rate model could be used to derive the sequence of discount factors $\left\{ v_t \right\}$ required for the recurrence relation in this paper because the relation simply requires that interest rates are derivable, and available when required for use in the calculations.  Notice that $\left\{ v_t \right\}$ is independent of the lives.

\subsubsection{Mortality}
     \label{Assumptions_Mortality}

The mortality table being used for a particular step must be known in advance of reaching the time step being simulated.  There is no need to assume that the underlying mortality cannot change, provided that it is possible to derive a set of survival probabilities from whatever mortality model is applicable throughout the period up to the point that the transition matrix is used.

There is a great body of work on the modelling of mortality rates; for example \cite{Cairns_Review} provide a comprehensive discussion on recent models.  Much of this work shows that mortality is currently improving over time, so that
\begin{quote}
  `the probability of a life currently aged 75 surviving for a year'
\end{quote}
is less than
\begin{quote}
  `the probability of a life aged currently 65 surviving from age 75 to age 76, assuming he first survives for 10 years to reach age 75'.
\end{quote}

While it is noted that these improvements in mortality exist, they are not of fundamental relevance to the workings of the recurrence relation derived in Section \ref{SurvivalProbabilities}.  Any mortality model could be used to derive the sequence of survival probabilities $\left\{ \mathbf{W}_{\mathbf{x},t,g} \right\}$ required for the recurrence relation in this paper because the relation simply requires that mortality rates are derivable, and available when required for use in the calculations.  It should also be noted that the granularity of the improvements in mortality might mean that the underlying table only needs to be changed every twelfth step, effectively using annual improvements in a projection which uses monthly steps.

\subsection{The Actual Algorithm}
     \label{ActualAlgorithm}
Explicitly outlining the algorithm highlights its computational complexity with respect to the number of steps for which the projection runs.  Sample timing results which support these complexities are given in Section \ref{Performance Improvement}.

\subsubsection{Existing Summation Algorithm}
     \label{ActualAlgorithm_Existing}
In order to calculate the reserve at each future time using the standard summation approach, which is a direct implementation of the formula in Equation \ref{basicReserveSummation}, two forward loops are required.  Let $S$ be the index of the maximum projection step, which may either be calculated from the data or set as a parameter.
Then the algorithm to calculate the reserves is
\begin{algorithmic}
\FOR{$t=0,1,\ldots{},S$}
    \STATE obtain $\overline{\mathbf{c}}_{\mathbf{x},t}, v_t, \overline{\mathbf{W}}_{\mathbf{x},t,f}, \overline{\mathbf{W}}_{\mathbf{x},t,1}$
\ENDFOR  
\FOR{$j$ in states}
    \STATE set $\overline{r}_{\mathbf{x},t,j}=0$
\ENDFOR  
\STATE set $v_\text{cash flow} = 1$
\STATE set $v_\text{end step} = 1$

\FOR{$t=0,1,\ldots{},S$}
    \FOR{$s=t,\ldots{},S$}
        \STATE set $v_\text{cash flow} = v_\text{end step} \times v_t^f$
        \STATE set $v_\text{end step}  = v_\text{end step} \times v_t$
        \FOR{$j$ in states}
            \STATE increment $\overline{r}_{\mathbf{x},t,j}$ by $v_\text{cash flow}
                              \times
                              \left(
                                      \displaystyle \prod_{r<t}{ \overline{\mathbf{W}}_{\mathbf{x},r,1} }
                              \right)
                              \times
                              \overline{\mathbf{W}}_{\mathbf{x},t,f}
                              \times \overline{\mathbf{c}}_{\mathbf{x},s}
                             $
        \ENDFOR  
    \ENDFOR  
\ENDFOR  
\end{algorithmic}
Hence, from the loop nest over $t$ and $s$, it is clear that obtaining the sequence $\{ \overline{\mathbf{r}}_{\mathbf{x},t} \}$ is $O(S^2)$, i.e. the computational complexity is quadratic in the number of projection steps.

\subsubsection{Proposed Recurrence Algorithm}
     \label{ActualAlgorithm_Proposed}
As stated in Section \ref{GeneralForm}, calculating the reserve at each future time using the vectorised recurrence approach allows the use of backward substitution, so that the nest of two forward loops is replaced by a single backward loop.  Again, let $S$ be the index of the maximum projection step.  Then the algorithm to calculate the reserves is
\begin{algorithmic}
\FOR{$t=0,1,\ldots{},S$}
    \STATE obtain $\overline{\mathbf{c}}_{\mathbf{x},t}, v_t, \overline{\mathbf{W}}_{\mathbf{x},t,f}, \overline{\mathbf{W}}_{\mathbf{x},t,1}$
\ENDFOR  
\FOR{$j$ in states}
    \STATE set $\overline{r}_{\mathbf{x},S+1,j}=0$
\ENDFOR  
\STATE calculate $v_t$ and $v_t^f$
    \FOR[descending]{$t=S,\ldots{},0$}
        \FOR{$j$ in states}
            \STATE set $\overline{\mathbf{r}}_{\mathbf{x},t}
                  = v_{t}^f \: \overline{\mathbf{W}}_{\mathbf{x},t,f} \: \overline{\mathbf{c}}_{\mathbf{x},t}
                  +  v_{t}  \: \overline{\mathbf{W}}_{\mathbf{x},t,1} \: \overline{\mathbf{r}}_{\mathbf{x+1},t}$
        \ENDFOR  
    \ENDFOR  
\end{algorithmic}
Hence, obtaining the sequence $\{ \overline{\mathbf{r}}_{\mathbf{x},t} \}$ is $O(S)$, i.e. the computational complexity is linear in the number of projection steps.

\subsection{Summary}
     \label{SummaryConfirmation}

Having derived the vector form of the recurrence using an arbitrary, unstructured, case the recurrence relation should hold for all non-unit linked policy types where the policy type may change (resulting from a change in the survival state of $\mathbf{x}$) at undetermined future times, so long as the nominal amount of each possible cash flow can be determined in advance.

It is important to notice that everything in this Section has been discussed in terms of unit timescales.  The derivation is independent of the scaling factor and therefore it applies directly to projection steps of any length, in any investigation, without any need for `rescaling'; the assumption in Section \ref{Assumptions_CashFlows} can therefore applied without further adjustments.

%% file: useCases.tex
\section{Use Cases}
  \label{UseCases}

Here we state the values with which the various components of Equation \ref{eq:generalForm} must be populated in order for the equation to be applicable to a small selection of contract types of interest.

\subsection{Single-Life Contracts}
     \label{SingleLifeContracts}

For a single life, the simplest model has only two states, alive and dead; for this model, the states to which it is possible to migrate can be tabulated as
  \begin{center}
    \begin{tabular}{c c l}
      Current State & $x_1$ & Possible Next State  \\
      \hline
      0 & alive & 0 1 \\
      1 & dead & 1 \\
    \end{tabular}
  \end{center}
which produces the transition matrix
\begin{align}
  \mathbf{W}_{\mathbf{x},t,g}  = \begin{pmatrix}  {}_{g}p_{x}  &  {}_{g}q_{x}  \\  0  &  1  \end{pmatrix}
  & &
  g \in \left\{ f,1 \right\}
  \label{Stochastic matrix for two state, single life model}
\end{align}
and hence the vector relation is
\begin{equation}
    \begin{pmatrix}  r_{\mathbf{x},t,0}  \\  r_{\mathbf{x},t,1}  \end{pmatrix}
      = v_{t}^f \: \begin{pmatrix}  {}_{f}p_{x}  &  {}_{f}q_{x}  \\  0  &  1  \end{pmatrix} \: \begin{pmatrix}  c_{\mathbf{x},t,0}  \\  c_{\mathbf{x},t,1}  \end{pmatrix}
      + v_{t} \: \begin{pmatrix}  {}_{1}p_{x}  &  {}_{1}q_{x}  \\  0  &  1  \end{pmatrix} \: \begin{pmatrix}  r_{\mathbf{x+1},t,0}  \\  r_{\mathbf{x+1},t,1}  \end{pmatrix}
  \label{Vector relation for two state, single life model}
\end{equation}

\subsubsection{Single Life Annuities}
        \label{SingleLifeAnnuities}
For single life annuities, Equation \ref{Vector relation for two state, single life model} can be intepreted directly as
\[
    \begin{pmatrix}  a^{\prime}_{x}  \\ 0  \end{pmatrix}
    = v^f \: \begin{pmatrix}  {}_{f}p_{x}  &  {}_{f}q_{x}  \\   0  &  1  \end{pmatrix} \: \begin{pmatrix}  c_{x,t}  \\ 0   \end{pmatrix}
    + v  \: \begin{pmatrix}  p_{x}  &  q_{x}  \\   0  &  1  \end{pmatrix} \: \begin{pmatrix}   a^{\prime}_{x+1}  \\ 0   \end{pmatrix}
\]
where the $j$ index on the cash flow has been dropped since there is only one non-trivial cash flow.  Removing the ZRS leads to
\[
    a^{\prime}_{x}
    = v^f \: {}_{f}p_{x} \: c_{x,t}
    + v   \:       p_{x} \: a^{\prime}_{x+1}
\]

Different values of $f$ lead to annuities where the timing differs, i.e. in advance, in arrear, or part-way through the step.  Varying $c_{\mathbf{x},t}$ leads to the recurrences for other differences in types of annuity;
\begin{itemize}
  \item[-] for a level annuity, $c_{x,t} = \theta $ where $\theta$ is constant,
  \item[-] for an increasing annuity, $c_{x,t} = \phi_t $ where $\phi_t$ increases in arithmetic progression,
  \item[-] for an escalating annuity, $c_{x,t} = \omega_t $ where $\omega_t$ increases in geometric progression,
  \item[-] for a limited term annuity, $c_{x,t} = 0 $ for all time steps after the end of the policy term.
\end{itemize}
Hence, for a level annuity of amount $\theta=1$, payable in advance (so that $f=0$), the relation is
\[
    \ddot{a}_{x} = 1 + v \: p_{x} \: \ddot{a}_{x+1}
\]
as presented in Section \ref{Single-LifeAnnuityCertain}.

For whole life annuities an appropriate boundary condition is ${}_{s}p_{120}=0$ for $s>0$, and for limited term annuities where the policy was effected by a life aged $x$ and the original term was $n$ an appropriate boundary condition is $a^{\prime}_{x+n:\lcRoof{0}}=0$.
Proceeding in this manner provides the annuity factor which applies to a life aged $x$ at time $t$.

\subsubsection{Single Life Endowments}
        \label{SingleLifeEndowments}

A pure endowment may be viewed as an annuity where all cash flows except one are zero, the non-zero cash flow being at the time the endowment is payable.  Using this interpretation, the usual value for the cash flow vector is $\mathbf{c}_{\mathbf{x},t} = \begin{pmatrix} 0 & 0 \end{pmatrix}^T$, and the non-zero cash flow for a policy effected by a life aged $x$ at inception with original term $n$ is $\mathbf{c}_{\mathbf{x+n},0} = \begin{pmatrix} 1 & 0 \end{pmatrix}^T$; this non-zero cash flow is also the boundary condition required to obtain the assurance factor as $r_{\mathbf{x},0}$ while $r_{\mathbf{x},1}$ is, again, redundant.
Therefore Equation \ref{Vector relation for two state, single life model} becomes
\[
    \begin{pmatrix}  \lcEnd{A}{x}{n}  \\ 0  \end{pmatrix}
    =
    v^f \:
    \begin{pmatrix}  {}_{f}p_{x}  &  {}_{f}q_{x}  \\   0  &  1  \end{pmatrix}
    \:
    \begin{pmatrix}  0  \\ 0   \end{pmatrix}
    +
    v  \:
    \begin{pmatrix}  p_{x}  &  q_{x}  \\   0  &  1  \end{pmatrix}
    \:
    \begin{pmatrix}   \lcEnd{A}{x+1}{n-1}  \\ 0   \end{pmatrix}
    \quad		\quad		\quad
\]
where $f \in [0,1]$ is arbitrary, and removing the ZRS leads to
\[
    \lcEnd{A}{x}{n} = v \: p_{x} \: \lcEnd{A}{x+1}{n-1}
\]
with the boundary condition $\lcEnd{A}{x+n}{0} = 1$.

\subsubsection{Single Life Assurances}
        \label{SingleLifeAssurances}

It is possible to show that assurances cannot use a Markov transition matrix with only two states. This is because the assumption that the cash flows are computable as a function of $x$, $t$ and state does not hold; this results from benefit being payable on the transition from one state to another, rather than the continuance in a particular state.

A straightforward solution is to introduce a third state, say `died in step', from which the life transfers to the dead state in the next step with probability 1.  Under this construction, ternary labelling must be used for the states, and the states to which it is possible to migrate can be tabulated as
  \begin{center}
    \begin{tabular}{c c l}
      Current State & $x_1$ & Possible Next State  \\
      \hline
      0 & alive & 0 1 \\
      1 & died in step & 2 \\
      2 & dead & 2 \\
    \end{tabular}
  \end{center}
which produces the transition matrix
\begin{align}
  \mathbf{W}_{\mathbf{x},t,g}  =
  \begin{pmatrix}
    {}_{g}p_x & {}_{g}q_x & 0 \\ 0 & 0 & 1 \\ 0 & 0 & 1
  \end{pmatrix}
  & &
  g \in \left\{ f,1 \right\}
  \label{Stochastic matrix for three state, single life model}
\end{align}
By setting the cash flow vector to $\mathbf{c}_{\mathbf{x},t} = \begin{pmatrix} 0 & 1 & 0 \end{pmatrix}^T$ for each step, the vector form of the equation becomes
\[
  \begin{pmatrix}  r_{\mathbf{x},t,0} \\ r_{\mathbf{x},t,1} \\ r_{\mathbf{x},t,2}  \end{pmatrix}
  = v^{f} \: \begin{pmatrix}  {}_{f}p_x & {}_{f}q_x & 0 \\ 0 & 0 & 1 \\ 0 & 0 & 1  \end{pmatrix}
      \: \begin{pmatrix}  0 \\ 1 \\ 0  \end{pmatrix}
  + v \: \begin{pmatrix}  p_x & q_x & 0 \\ 0 & 0 & 1 \\ 0 & 0 & 1  \end{pmatrix}
      \: \begin{pmatrix}  r_{\mathbf{x+1},t,0} \\ r_{\mathbf{x+1},t,1} \\ r_{\mathbf{x+1},t,2}  \end{pmatrix}
\]

Even though another state has been introduced, the boundary conditions used in Section \ref{SingleLifeAnnuities} are still appropriate; i.e. ${}_{s}p_{120}=0$ for $s>0$ in the case of a whole life assurance, and $r_{\mathbf{x+n},0} = 0$ in the case of a term assurance.
Removing the ZRSs leads to
\[
  r_{x,t,0} = v^{f} \: {}_{f}q_x + v \: p_x \: r_{x+1,t,0}
\]
By setting $f=1$, it is apparent that, depending on the boundary condition used, this is equivalent to
either a single-life whole life assurance, so that the relation is
\[
  A_{x} = v \: q_{x} +v \: p_{x} \: A_{x+1}
\]
or a single-life term assurance, where the relation is
\[
  \lcTerm{A}{x}{n} = v \: q_{x} +v \: p_{x} \: \lcTerm{A}{x+1}{n-1}
\]

\subsection{Two-Life Contracts}
     \label{TwoLifeContracts}

For two lives, using $x$ and $y$ rather than $x_1$ and $x_2$, the states to which it is possible to migrate can be tabulated as
  \begin{center}
    \begin{tabular}{c c c c l}
      Current State & Binary & $x$ & $y$ & Possible Future States \\
      \hline
      0 & 00 & alive & alive & 0 1 2 3 \\
      1 & 01 & alive & dead  & 1 3 \\
      2 & 10 & dead  & alive & 2 3 \\
      3 & 11 & dead  & dead  & 3 \\
    \end{tabular}
  \end{center}
and the complete transition matrix is
\begin{align*}
  \mathbf{W}_{\mathbf{x},t,g}
  =\begin{pmatrix}
     {}_{g}p_{x} \: {}_{g}p_{y}  &  {}_{g}p_{x} \: {}_{g}q_{y}  &  {}_{g}q_{x} \: {}_{g}p_{y}  &  {}_{g}q_{x} \: {}_{g}q_{y}
        \\
     0  &  {}_{g}p_{x}  &  0  &  {}_{g}q_{x}
        \\
     0  &  0  &  {}_{g}p_{y}  &  {}_{g}q_{y}
        \\
     0  &  0  &  0  &  1
   \end{pmatrix}
  & &
  g \in \left\{ f,1 \right\}
\end{align*}

\subsubsection{Two-Life Annuities}
        \label{TwoLifeAnnuities}

Many practising actuaries are only interested in a) single life annuities and b) reversionary annuities (where the benefit is payable to a second life after the death of the first, so long as the survivorship follows the ordering specified in the policy document).  However, there are other forms of two-life annuity, i.e. a) joint-life annuities (which are payable so long as both lives are alive at the time of payment), and b) last survivor annuities (which are payable so long as at least one of the lives are alive at the time of payment).

As with single life annuities, different values of $f$ lead to annuities where the timing differs, and varying $\mathbf{c}_{\mathbf{x},t}$ leads to the relations for differences in types of annuity.  However, in contrast to single life annuities, the recurrence equation has different interpretations depending on how the cash flow vector is populated;
\begin{enumerate}[a)]
  \item  when $\mathbf{c}_{\mathbf{x},t} = \begin{pmatrix} 0 & 0 & 1 & 0 \end{pmatrix} ^T$, $\mathbf{r}_{\mathbf{x},t}$ can be interpreted as $\begin{pmatrix}  a^{\prime}_{x|y} & 0 & a^{\prime}_{y} & 0  \end{pmatrix} ^T$ which is precisely what is required to obtain the reserve factors for a reversionary annuity, per the derivation in Section \ref{SurvivalBasedPresentation}.

  \item  when $\mathbf{c}_{\mathbf{x},t} = \begin{pmatrix} 1 & 0 & 0 & 0 \end{pmatrix} ^T$, $\mathbf{r}_{\mathbf{x},t}$ can be interpreted as $\begin{pmatrix}  a^{\prime}_{x,y} & 0 & 0 & 0  \end{pmatrix} ^T$ which corresponds to the recurrence required to obtain for the reserve factors for a joint life annuity.

  \item  when $\mathbf{c}_{\mathbf{x},t} = \begin{pmatrix} 1 & 1 & 1 & 0 \end{pmatrix} ^T$, $\mathbf{r}_{\mathbf{x},t}$ can be interpreted as $\begin{pmatrix} a^{\prime(LS)}_{x,y} & a^{\prime}_{x} & a^{\prime}_{y} & 0  \end{pmatrix} ^T$ which corresponds to the recurrence required to obtain for the reserve factors for a last survivor annuity.  The probability that a payment on a last survivor annuity is made is usually given as $1-{}_{f}q_{x} \: {}_{f}q_{y}$, i.e. the probability that both are not dead; however, it is straightforward to show that
         \[
              1-{}_{f}q_{x} \: {}_{f}q_{y} 
                                            = {}_{f}p_{x} \: {}_{f}p_{y} + {}_{f}p_{x} \: {}_{f}q_{y} + {}_{f}q_{x} \: {}_{f}p_{y}
          \]
         so that the top row of $\mathbf{W}_{\mathbf{x},t,f}$ is, indeed, the required probability for payment on a last survivor annuity when the cash flow vector is populated as given.
\end{enumerate}

For whole life annuities which are payable on two lives, appropriate boundary conditions are ${}_{s}p_{120}=0$ for $s>0$.  For limited term policies, an appropriate condition $a^*=0$ where $a^*$ is the reserve factor at maturity of the policy.

\subsubsection{Two-Life Endowments}
        \label{TwoLifeEndowments}

Section \ref{SingleLifeEndowments} uses the observation that a single life endowment can be regarded as an annuity with only one non-zero payment; a similar interpretation can be applied to two-life endowments.  By setting all cash flows except the one at maturity to zero, reserve factors for two-life endowments can be obtained by setting $\mathbf{c}_{\mathbf{x},t} = \begin{pmatrix} 0 & 0 & 0 & 0 \end{pmatrix} ^T$ in all steps and using $\mathbf{c}_{\mathbf{x+n},0} = \begin{pmatrix} 1 & 0 & 0 & 0 \end{pmatrix} ^T$ to give $\lcEnd{A}{x,y}{n}$ as the first element of $\mathbf{r}_{\mathbf{x},t}$, i.e. having removed ZRSs, the relation in standard notation is
\[
    \lcEnd{A}{x,y}{n} = v \: {}_{g}p_{x} \: {}_{g}p_{y}\: \lcEnd{A}{x+1,y+1}{n-1}
\]
subject to the boundary condition
\[
    \lcEnd{A}{x+n,y+n}{0} = 1
\]

\subsubsection{Two-Life Assurances}
     \label{TwoLifeAssurances}

For two-life assurances, the third state (died in step) must be applied to both lives and so the transition matrix for this model is
\begin{align*}
  \mathbf{W}_{\mathbf{x},t,g} =
   \begin{pmatrix}
    {}_{g}p_x \: {}_{g}p_y & {}_{g}p_x \: {}_{g}q_y & 0 & {}_{g}q_x \: {}_{g}p_y & {}_{g}q_x \: {}_{g}q_y & 0 & 0 & 0 & 0 \\
    0 & 0 & {}_{g}p_x & 0 & 0 & {}_{g}q_x & 0 & 0 & 0 \\
    0 & 0 & {}_{g}p_x & 0 & 0 & {}_{g}q_x & 0 & 0 & 0 \\
    0 & 0 & 0 & 0 & 0 & 0 & {}_{g}p_y & {}_{g}q_y & 0 \\
    0 & 0 & 0 & 0 & 0 & 0 & 0 & 0 & 1 \\
    0 & 0 & 0 & 0 & 0 & 0 & 0 & 0 & 1 \\
    0 & 0 & 0 & 0 & 0 & 0 & {}_{g}p_y & {}_{g}q_y & 0 \\
    0 & 0 & 0 & 0 & 0 & 0 & 0 & 0 & 1 \\
    0 & 0 & 0 & 0 & 0 & 0 & 0 & 0 & 1 \\
   \end{pmatrix}
  & &
  g \in \left\{ f,1 \right\}
\end{align*}

As with other policy types described above, keeping $\mathbf{c}_{\mathbf{x},t}$ fixed leads to level two-life assurances, whereas having a varying $\mathbf{c}_{\mathbf{x},t}$ leads to increasing assurances, escalating assurances, or fixed-term assurances, depending on the flavour of the variation.

When the benefit is payable on the first death, cash flow vectors for level assurances should be
\[
  \mathbf{c}_{\mathbf{x},t} = \begin{pmatrix} 0 & 1 & 0 & 1 & 1 & 0 & 0 & 0 & 0 \end{pmatrix}^T
\]
and when the benefit is payable on the second death, the cash flow vectors for level assurances should be
\[
  \mathbf{c}_{\mathbf{x},t} = \begin{pmatrix} 0 & 0 & 0 & 0 & 1 & 1 & 0 & 1 & 0 \end{pmatrix}^T
\]

\subsection{More Than Two Lives}
     \label{More Than Two Lives}
Although cases where there are more than two lives are uncommon, we include them for completeness.

The two-life models in Section \ref{TwoLifeContracts} are straightforward extensions of the single life models in Section \ref{SingleLifeContracts}.  For larger numbers of lives, the transition matrices may be obtained by induction, using tensor products.

Let $\mathbf{W}_{\mathbf{x},t,g}^m$ be the transition matrix required for $m$ lives. Then
\begin{enumerate}[a)]
  \item  for the two state model (used for annuities and endowments),
         \[
           \mathbf{W}_{\mathbf{x},t,g}^m = \mathbf{W}_{\mathbf{x},t,g}^{m-1}  \otimes
           		\begin{pmatrix}  {}_{g}p_{x_m}  &  {}_{g}q_{x_m}  \\   0  &  1  \end{pmatrix}
         \]
         with $\mathbf{W}_{\mathbf{x},t,g}^0=(1)$; there are $2^m$ possible states so that $\mathbf{W}_{\mathbf{x},t,g}^m$ has $4^m$ elements, only $3^m$ of which are non-zero.
  \item  for the three state model (used for assurances),
         \[
           \mathbf{W}_{\mathbf{x},t,g}^m = \mathbf{W}_{\mathbf{x},t,g}^{m-1}  \otimes
           		\begin{pmatrix}  {}_{g}p_{x_m} & {}_{g}q_{x_m} & 0 \\  0 & 0 & 1 \\  0 & 0 & 1  \end{pmatrix}
         \]
         with $\mathbf{W}_{\mathbf{x},t,g}^0=(1)$; there are $3^m$ possible states so that $\mathbf{W}_{\mathbf{x},t,g}^m$ has $9^m$ elements, only $4^m$ of which are non-zero.
\end{enumerate}

\subsection{With-Profit Policies}
     \label{With-ProfitPolicies}
The algorithm can be applied to with-profit policies so long as the cash flows can be determined in advance of their use so that the assumption in Section \ref{Assumptions_CashFlows} holds.  This should not pose any practical problems because the cash flows can be calculated in the forward loop and then used in the backward loop.  Even if the bonuses are applied on a two-tier basis, this algorithm can be used if estimated bonus rates are available in advance of calculating the bonuses attaching at a particular time.

\subsection{Other Benefits}
     \label{OtherBenefits}
The algorithm can also be applied to other types of policies; for example Critical Illness, Permanent Health Insurance and Unemployment Income policies may all be modelled using a Markov transition matrix derived from the relevant state transition diagram.
Given the reversible nature of some of the transitions (say, between the `able' and `ill' states of a P.H.I. policy), the Markov transition matrix might not be upper triangular.

\subsection{Step Lengths}
     \label{StepLengths}

In all of the cases considered in Sections \ref{SingleLifeContracts} and \ref{TwoLifeContracts}, the relationships are based on time steps of unit length; in theoretical work, it is usually assumed that the default length of a step is a year.  Cases where cash flows happen more frequently need to be adjusted to allow for the frequency of payments.

Using this vector form of the recurrence there is no need for such adjustments; the sequence $\left \lbrace \mathbf{c}_{\mathbf{x},t} \right \rbrace$ indicates the nominal amount of each cash flow.  Hence, in the case of escalating payments, for example, the stream of payments which populate the sequence $\left \lbrace \mathbf{c}_{\mathbf{x},t} \right \rbrace$ must already include the allowance for escalation and the timing of the cash flows.  Similarly, if a projection is being performed using monthly steps and the cash flows occur yearly, then the sequence $\left \lbrace \mathbf{c}_{\mathbf{x},t} \right \rbrace$ would have a non-zero value for every twelfth step only.

\subsection{Summary}
     \label{SummaryVectors}

The vector form of the recurrence automatically allows for all possible combinations of lengths of time steps, and variability in cash flow, simply by populating the sequence of cash flow vectors appropriately, and populating $\mathbf{W}_{\mathbf{x},t,g}$ accordingly.  For each of the two-state and three-state models presented here, there is only one transition matrix for $m$ lives; how each matrix is applied will vary depending on the type of policy under consideration.

The examples presented in this section have an equivalent interpretation to relationships which may be derived from first principles using the relevant summation for a particular type of policy; the derivations are straightforward, if laborious.

The collection of recurrences described above indicates that this vector approach can be applied to a wide range of contracts; the collection of policy types included here is not intended to be exhaustive, but it is meant to demonstrate that any potential restrictions which appear to apply to non-unit linked policies can be overcome, and that the relation can be used in the form presented here.

%% file: performance.tex
\section{Performance Implications}
  \label{Performance Implications}

This paper is mainly theoretical, and is intended as the basis for future implementations; the timing results included here illustrate the potential performance benefits of our recurrence relation approach.

To test the performance of our implementation, we created some synthetic data which has properties which are representative of a cohort of people who retired recently, either at normal retirement age, or slightly early.  The main characteristics of these data are; the date of birth is uniformly distributed so that the age at the valuation date is between 57 and 67; the policy inception date is uniformly distributed over the calendar year prior to the valuation date, so that these policies represent a cohort of new business; roughly 75\% of the policyholders are male; roughly 80\% of the policies have payments made monthly, the remainder having payments made annually; the amount paid at each payment has a log-normal distribution with a mean of roughly 5.0 and standard deviation of about 1.5; approximately 95\% of the policies have no escalation and the remainder escalate at either 3\%, 4.25\% or 5\% on each policy anniversary.  For two-life policies, both lives were drawn from the same uniform distribution of ages, with no regard as to which of the lives was older.

All timings were obtained using a laptop with an Intel i5-2450 2.5GHz dual core CPU running {\it Windows} 7, and using Intel {\it ifort} 12.1.  Although this platform is primarily used for development, it gives an indication of performance available within industry where PCs with i5 and i7 CPUs are widely available.

\subsection{Effect of Changing the Algorithm}
     \label{Effect of Changing the Algorithm}

\subsubsection{Measuring Performance}
        \label{Measuring Performance}

Our starting point was a Fortran 90 code which closely mirrored the code produced by one of the commercial actuarial valuation packages in that a) the financial results produced by our program agreed to those from the commercial system to the 7 significant figures output when the same demographic and economic assumptions were used, and b) the performance of our original code was almost identical to that obtainable from the commercial package.  The code went through several stages of optimisation including implementation of the vector version of the recurrence approach; this was not particularly difficult, and led to significant speedup in its own right.  The code was then parallelised using OpenMP, with a decomposition of policies across available threads, and a dynamic loop schedule to overcome any potential load imbalance resulting from different policies requiring different numbers of time steps.  A more complete discussion of that optimisation is given in our previous work \citep{TuckerBull_WHPCF}.

\subsubsection{Performance Improvement}
        \label{Performance Improvement}

Figure \ref{scaling_algorithmChange} shows the average time to process single-life annuity policies using the two algorithms in Section \ref{ActualAlgorithm}; the implementations of the algorithms had the same level of optimisation, so that the only difference between the runs was the algorithm.  It is clear that the shapes of the curves in that graph match the complexities stated in Sections \ref{ActualAlgorithm_Existing} and \ref{ActualAlgorithm_Proposed}.

Also, on the scale presented to demonstrate the quadratic nature of the summation approach, the times for the recurrence approach are almost indistinguishable from the axis; this indicates that as well as having better scaling than the summation approach, the recurrence approach is far faster than the summation approach; in fact, recursion is two about orders of magnitude faster than summation.

\subsection{Scalability of the Algorithm}
     \label{Scalability of the Algorithm}

A highly optimised {\it ab-initio} implementation of our recurrence relation approach was produced in Fortran 90, and its performance was measured for annuity contracts.

\subsubsection{Timing Methodology}
        \label{Timing Methodology}
To allow for operating system daemons, each timing run was performed seven times; the minimum of those runs is reported.  This method produces a smooth progression because the times reported approach those which are the best which might be obtained if nothing else was running on the system.

\subsubsection{Performance Results}
        \label{Performance Results}

Figure \ref{scaling_annuities} shows the average time to process single-life and reversionary annuity policies using the the recurrence approach to calculating reserves; it is clear that using the recurrence approach creates linear scaling with number of steps for both types of annuity; this agrees with the complexities derived in Section \ref{ActualAlgorithm_Proposed}.  The fact that the slope of the line for the reversionary annuity policies is roughly twice the slope for the single-life annuity policies is to be expected because, for a reversionary annuity, there are twice as many lives for which $\left \{ l_\mathbf{x} \right \}$ needs to be obtained and, at each step, two reserve factors are required, per Equation \ref{eq:revAnnRecRelVec}.

\subsection{Predicted Performance}
     \label{Predicted Performance}

\subsubsection{Representative Portfolio}
        \label{Representative Portfolio}

The information obtained from placing timers in our optimised implementation of the recurrence algorithm is shown in Table \ref{time_400000_policies_laptop}.

The fact that calculating $\left \{ l_\mathbf{x+t} \right \}$ takes twice as long for two-life policies as for single life policies is to be expected since, on average, twice as many $l_{x}$ calculations are required, and this is simply a result of the fact that there twice as many lives, and in our synthetic data they are drawn from the same uniform distribution of ages at the valuation date.

\subsubsection{Predicted Results}
        \label{Predicted Results}

Table \ref{representative_portfolio_500000} shows the composition of a representative portfolio of 500,000 annuity policies.

Using the results from Table \ref{time_400000_policies_laptop}, the wall-clock time to process that representative portfolio using both cores of the dual core laptop is estimated to be 5.7 seconds. Since the ages at the valuation date have a uniform distribution, the average age (at the valuation date) of the representative portfolio is 62.  Therefore, assuming a limiting age of 120, the average outstanding projection term is 58 years  (i.e. 696 months).

Hence, using the same reasoning as in Section \ref{BruteForceApproach}, the time to process a `brute-force' solvency calculation, using 1000 scenarios at each future monthly step, is estimated to be
\[
  1000 \times \left[ \frac{696}{696} + \frac{695}{696} + \frac{694}{696} + \cdots{} + \frac{1}{696} \right]
  \times 5.7
  \text{ sec }
  \approx 550 \text{ hr.}
\]
This is the time estimate using one dual-core laptop; clearly, it compares extremely favourably with the estimate of about 16 years obtained in Section \ref{BruteForceApproach}.  However, Section \ref{FutureWork} discusses the possibility that this time could still be significantly reduced by using different hardware.

%% file: conclusion.tex
\section{Conclusion}
  \label{Conclusion}

\subsection{Overview}
     \label{Overview}

This work has demonstrated that a vector form of recurrence relations can be found for many non-unit linked life assurance contracts of arbitrary complexity, and has provided specific recurrence relations for several exemplar contracts.  It has demonstrated that efficient implementations of the recurrence relations have scalability which is particularly favourable.

\subsection{Future Work}
     \label{FutureWork}

As part of our investigation into the use of High Performance Computing techniques, this work forms the foundation of an implementation of the brute force approach to calculating liabilities for use in demonstrating solvency.  Work to transfer vectorised code to a 48-core SMP, using OpenMP to coordinate the threads, has already begun.  It is expected that the scalability of this method is particularly good; given the results of our previous work, we expect to achieve over 95\% efficiency on the highest numbers of cores.

Following that, we plan to implement a version of the calculations on Graphics Processing Units (GPUs). GPUs are ideal for problems which require a large amount of processing on a relatively small amount of data, thereby overcoming the cost of transferring the data to the device.  For a projection of $N$ policies over $S$ steps, the amount of data required is $O(N)$, and the amount of computation (using a recurrence approach) is $O(NS)$, making these solvency calculations especially well-suited to such an implementation.

GPUs can process data roughly 50 to 60 times more quickly than CPUs; for the representative portfolio in Section \ref{Predicted Performance}, a single GPU might be able to process 1000 scenarios at each future step in roughly 10 hours, or `over night', and a cluster of, say, 20 GPUs would be able to do the processing `over lunch'.  This brings a full brute force calculation of liabilities for the solvency calculation within the realms of practicality.

%% file: acknowledge.tex
\section*{Acknowledgements}

We would like to thank Gavin Conn for his comments on drafts of Sections \ref{Motivation}, \ref{Derivation} and \ref{UseCases} of this paper.

%% file: representativePortfolio.tex
\begin{table}
\begin{center}
  \begin{tabular}{l c c c c}
    Annuity Type  &  Single  &  Joint  &  Reversion  &   Last   \\
          ~       &   Life   &  Life   &      ~      &  Survivor  \\
    \hline
    Number of lives  &  1  &  2  &  2  &  2  \\
    Number of reserve factors  &  1  &  1  &  2  &  3  \\
    \hline
    Time to calculate $\lbrace l_\mathbf{x+t} \rbrace$  & 2.67 & 5.32 & 5.32 & 5.19  \\
    Time to calculate the reserve factors  & 2.54 & 3.18 & 5.11 & 8.77  \\
    \hline
    Total processing time  & 6.73 & 10.39 & 12.37 & 15.90  \\
  \end{tabular}
  \caption{Time (seconds) to process 400,000 annuities of each type (using 2 threads on both cores of an Intel i5-2450 2.5GHz dual core CPU).
           Note that these are `total computing' times, so that wall-clock times would be about 50\% of the figures given.
          }
  \label{time_400000_policies_laptop}
\end{center}
\end{table}

%% file: runtimes.tex
\begin{table}
\begin{center}
  \begin{tabular}{l c c c c}
    Policy Type  &  Single  &  Joint  &  Reversion  &   Last   \\
          ~      &   Life   &  Life   &      ~      &  Survivor  \\
    \hline
    Number of policies  & 300,000 & 50,000  & 100,000  & 50,000   \\
  \end{tabular}
  \caption{Composition of a representative portfolio of 500,000 annuity policies.}
  \label{representative_portfolio_500000}
\end{center}
\end{table}

%% file: plot1.tex
\begin{figure}
  \begin{center}
    \includegraphics[  width=140mm      
                    , height=175mm      
                    ]{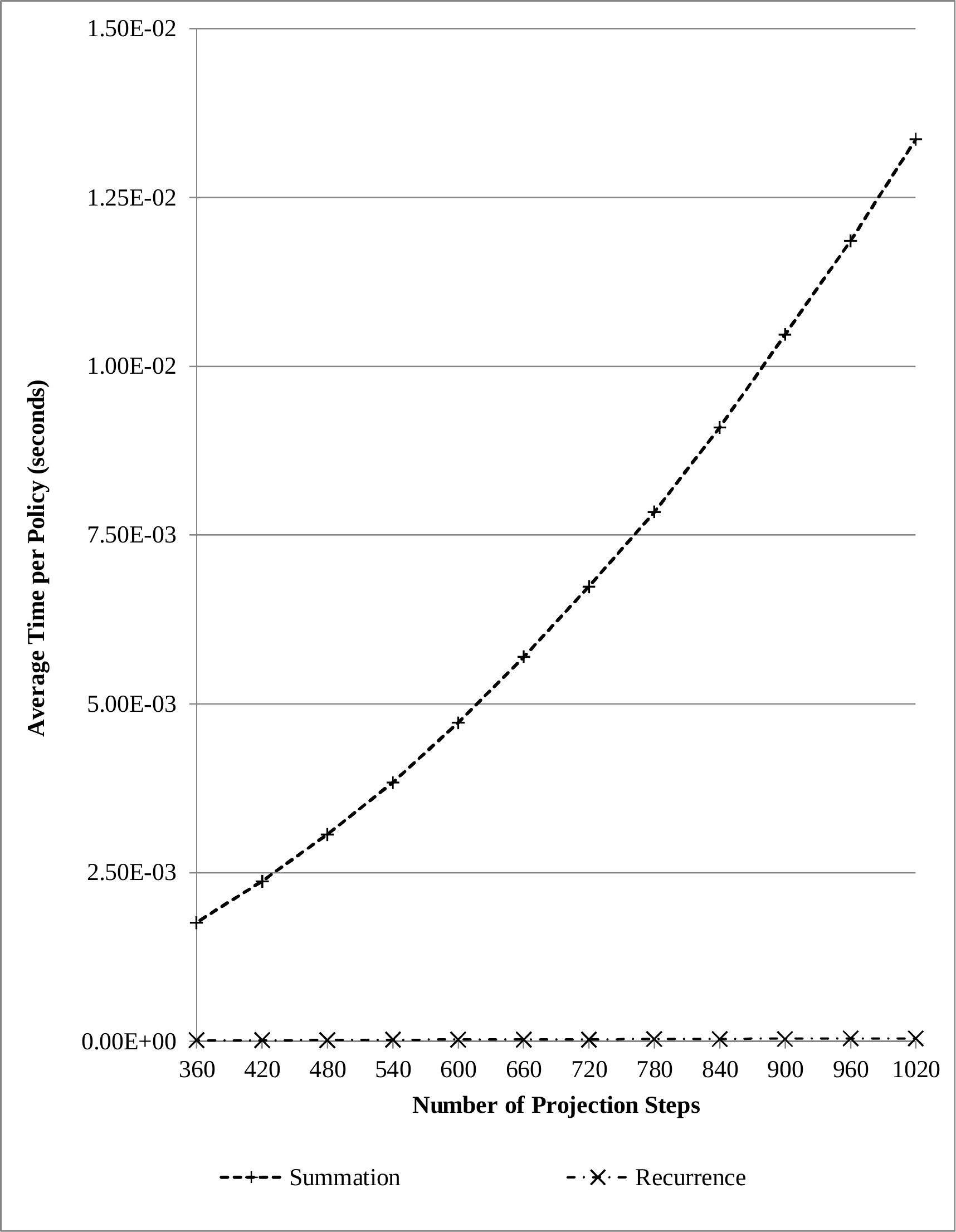}
    \caption{Time to calculate single life annuity reserves, averaged over 60,000 policies, using naive summation and recurrence algorithms on an Intel i5-2450 2.5GHz dual core CPU.}
    \label{scaling_algorithmChange}
  \end{center}
\end{figure}

%% file: plot2.tex
\begin{figure}
  \begin{center}
    \includegraphics[  width=140mm      
                    , height=174mm      
                    ]{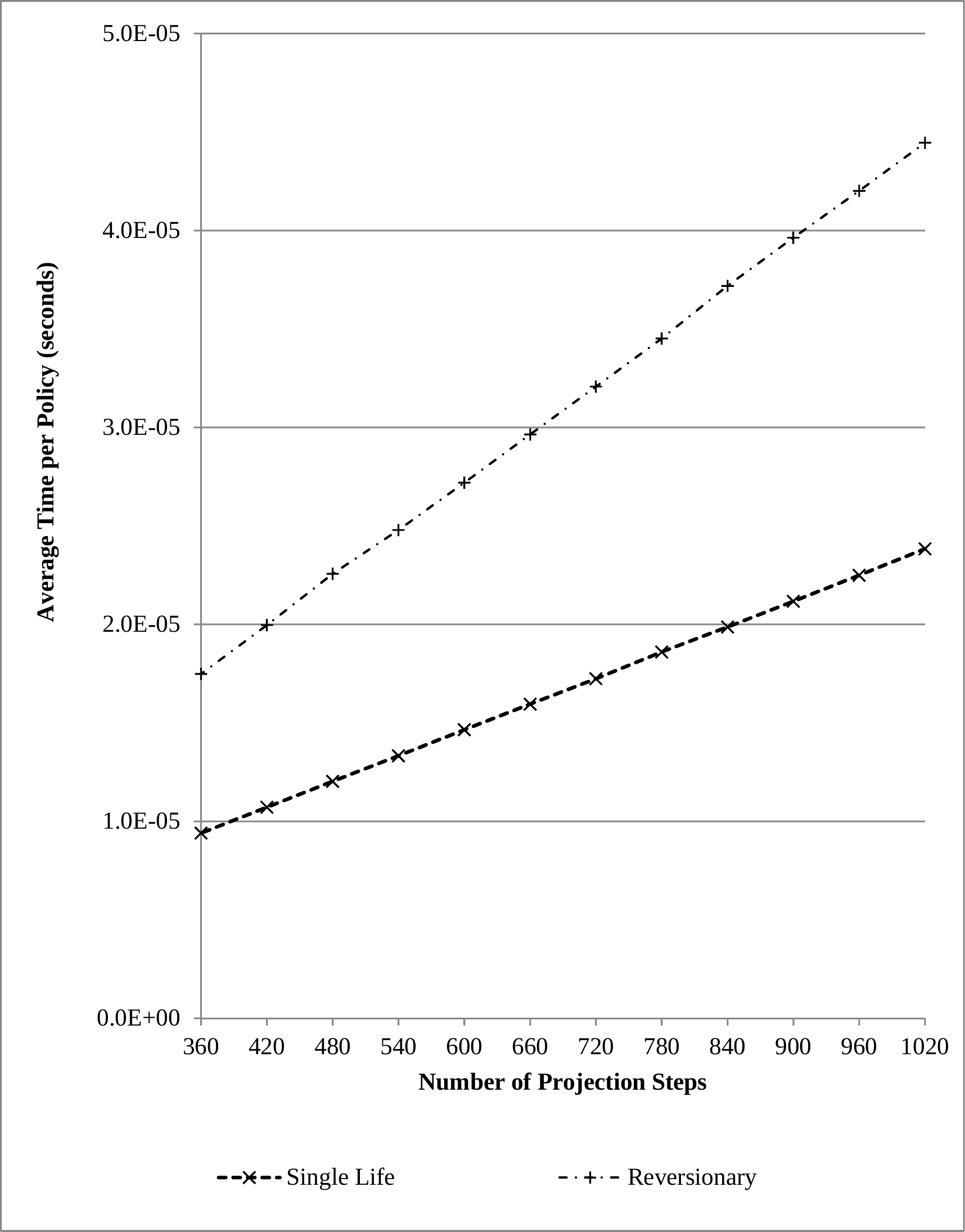}
    \caption{Time to calculate single life and reversionary annuity reserves, averaged over 60,000 policies, using recurrence algorithms on an Intel i5-2450 2.5GHz dual core CPU.}
    \label{scaling_annuities}
  \end{center}
\end{figure}